\documentclass[12pt,a4paper]{article}
\usepackage[T2A]{fontenc}
\usepackage[utf8]{inputenc}
\usepackage[english,russian]{babel}
\usepackage{graphicx}
\usepackage{aas_macros}

\usepackage{xcolor}
\usepackage{indentfirst}
\usepackage{cite}

\def\edm{\color{black}}

\title{On Systematic Errors in Cross-Identification of Binary Stars in Gaia}

\begin{document}

\author{E.D.~Malik, P.V.~Kaigorodov, D.A.~Kovaleva, O.Yu.~Malkov}

\maketitle

\begin{abstract}
{Using Gaia DR3 data, binary star catalogs have been created containing information on a total of more than {\edm 2.6 million} pairs. This increases by more than an order of magnitude the ensemble of binary stars with known characteristics, which previously numbered about {\edm 140 thousand pairs.} To perform statistical analysis of the complete ensemble of binary stars, including both previously known and newly discovered pairs, cross-identification by coordinates was carried out between the most complete pre-Gaia publication {\edm compilative} binary star catalog ILB and data from binary star catalogs based on Gaia DR3 results. An analysis of the results of this identification was performed, showing the dependence of its characteristics both on the data from the source catalogs and on coordinates. It is shown that in dense stellar fields, particularly in the Galactic disk, an increase in the fraction of false positive identifications can be expected. At the same time, for systems with large proper motion, there is a high probability of a false negative outcome. Possible modifications to the identification method are proposed to reduce the role of the described systematic errors and increase the reliability of its results.}    
\end{abstract}

\section{Introduction}
The comprehensive study of binary and multiple stellar systems is associated with a number of difficulties. In particular, different catalogs and identification systems are created for different observational types of multiple systems. To verify whether a star is binary, a researcher is forced to manually search for information about it in a large number of different sources, while facing the heterogeneity of existing identification systems and data. Special problems may arise when it is necessary to check the multiplicity of a large number of stars at once for statistical analysis, etc.

To solve these and a number of other problems, it is necessary to have a resource that combines data on binary and multiple systems of different (preferably all) observational classes.
While for specific observational classes of binary stars such resources have existed for quite some time {\cite{2001AJ....122.3480H, 2001AJ....122.3472H, 2002A&A...384..180F, 2002yCat.1274....0D, 2004A&A...424..727P,
2006yCat.5123....0D, 2006A&A...455.1165L, 2007A&A...469..807L, 2011yCat....102018R, 2012A&A...546A..69M, 2013AN....334..860A,
2014MNRAS.444.1982A, 2015BaltA..24..395M, 2016yCat....102026M, 2016yCat....102034M, 2017ARep...61...80S}}
or were created recently \cite{2020MNRAS.491.5489M, 2022MNRAS.517.2925C, 2023ARep...67..576Z, 2025Ap&SS.370....1Z},
a unifying resource was absent for a long time,
until the development of the Binary star Data Base (BDB) / Base de Donn\'ees
des etoiles Binaires 
began at the Besançon Observatory
\cite{2004ASPC..314..217O, 2009ASPC..411..442M, 2011AIPC.1346..134M}, which was later transferred
to the Institute of Astronomy of the Russian Academy of Sciences \cite{2012BaltA..21..309K, 2015A&C....11..119K, 2019ASPC..521..217K}.

BDB continues to be supplemented with catalogs of binary and multiple stars of various observational classes
\cite{2015BaltA..24..446K, 2017ASPC..510..360M}
and currently contains data on physical and positional parameters of 3859847 components of 2452572 systems
\cite{2023OAst...32..215K}.
Simultaneously, a number of technical problems related to correct identification
and {\edm cross-identification}
of BDB objects are being solved.
Thus, a system for identifying
components, pairs, and systems of multiple objects BSDB was developed and approved by the International Astronomical Union \cite{2015BaltA..24..185K},
and a master catalog of the database was created (and is constantly maintained),
the Identification List of Binaries (ILB) \cite{2016BaltA..25...49M}.
These efforts are aimed at facilitating the connection to BDB of both new catalogs of binary systems
\cite{2019RAA....19...33K},
and, if necessary, objects from modern large sky surveys.

Using Gaia DR3 data \cite{2023A&A...674A...1G}, binary star catalogs have been created containing information on a total of more than {\edm 2.4} million pairs. This significantly increases the ensemble of binary stars with known characteristics, which previously numbered no more than {\edm 144845} pairs. To perform statistical analysis of the complete ensemble of binary stars, it is necessary to combine data on previously known and newly discovered pairs.
In a previous work, primary cross-identification of ILB with data from binary star catalogs based on Gaia DR3 results was performed\cite{Malik2024}.

{\edm Cross-identification} of objects in stellar catalogs is a non-trivial task; practically always, some identifications may be incorrect, since the choice of {\edm cross-identification} criteria cannot account for all special cases with a large number of objects. Detection of {\edm cross-identification} errors is difficult and most often carried out manually, for some selected objects that, for some reason, attract researchers' attention. A systematic analysis of the results of cross-identification of Gaia project results with other surveys \cite{2019A&A...621A.144M} demonstrated the dependence of results on the angular resolution of catalogs {\edm (the accuracy of coordinates specified in the catalogs)} and the importance of choosing the identification radius in accordance with this characteristic. In the case of cross-identification of a compilative catalog, such as ILB, the angular resolution of the catalog as a whole cannot be determined. Within the framework of this work, we attempt to qualitatively assess the reliability of our previously performed {\edm cross-identification} and determine directions for possible improvement in the reliability of source identification in heterogeneous astronomical catalogs.

\section{Data Used}\label{data_used}

Cross-identification was carried out between ILB and catalogs based on Gaia mission results. Non-single star catalogs were published as part of Gaia DR3 (Non-Single Stars, hereinafter NSS) \cite{2023A&A...674A..34G}. These catalogs contain data on Gaia DR3 sources whose astrometric, photometric, or spectroscopic data indicate the presence of unresolved binarity (in some cases, multiplicity) or an invisible companion. In all cases, binary stars from NSS catalogs are represented in Gaia as one light source and have one Gaia identifier. The NSS catalogs contain more than 800,000 such sources. In this work, we will consider all such sources to be binary stars.

The wide binary star catalog ("Wide Binaries from Gaia EDR3", WB) was created as a result of searching Gaia EDR3 for pairs of stars with common proper motion located at a distance of no more than 1~pc from each other \cite{2021MNRAS.506.2269E}.
Thus, binary stars in the WB catalog have in all cases two Gaia identifiers for two components. Of the wide binary stars in the WB catalog, more than 1 million pairs are "reliable", i.e., with high probability they are physically bound.

\section{Results of Cross-Identification by Coordinates}

In \cite{Malik2024}, we presented the results of cross-identification of stars from Gaia catalogs and data included in the {\edm compilative} ILB catalog. In this work, we investigate the discrepancies found during {\edm cross-identification} between the coordinates of stars in ILB and Gaia. It is important to note that the results presented below were obtained under the following condition: in the case of impossibility to unambiguously identify a star from ILB with one or more Gaia stars due to a large number of objects falling within the identification circle, none of these objects were taken into consideration to avoid introducing errors into the relationships under consideration.

Discrepancies when comparing coordinates of the same star in two different catalogs are possible for several reasons: due to proper motion of stars, inaccuracy of the compared catalogs, or possible systematic errors. If errors are distributed non-uniformly across the celestial sphere, analysis of the distribution can provide information about systematic error either in the source data or in the {\edm cross-identification} methods.
In this work, we investigate the distance between identified objects—entities in ILB and Gaia identified as one object.

\begin{figure}
    \centering
    \includegraphics[width=0.49\textwidth]{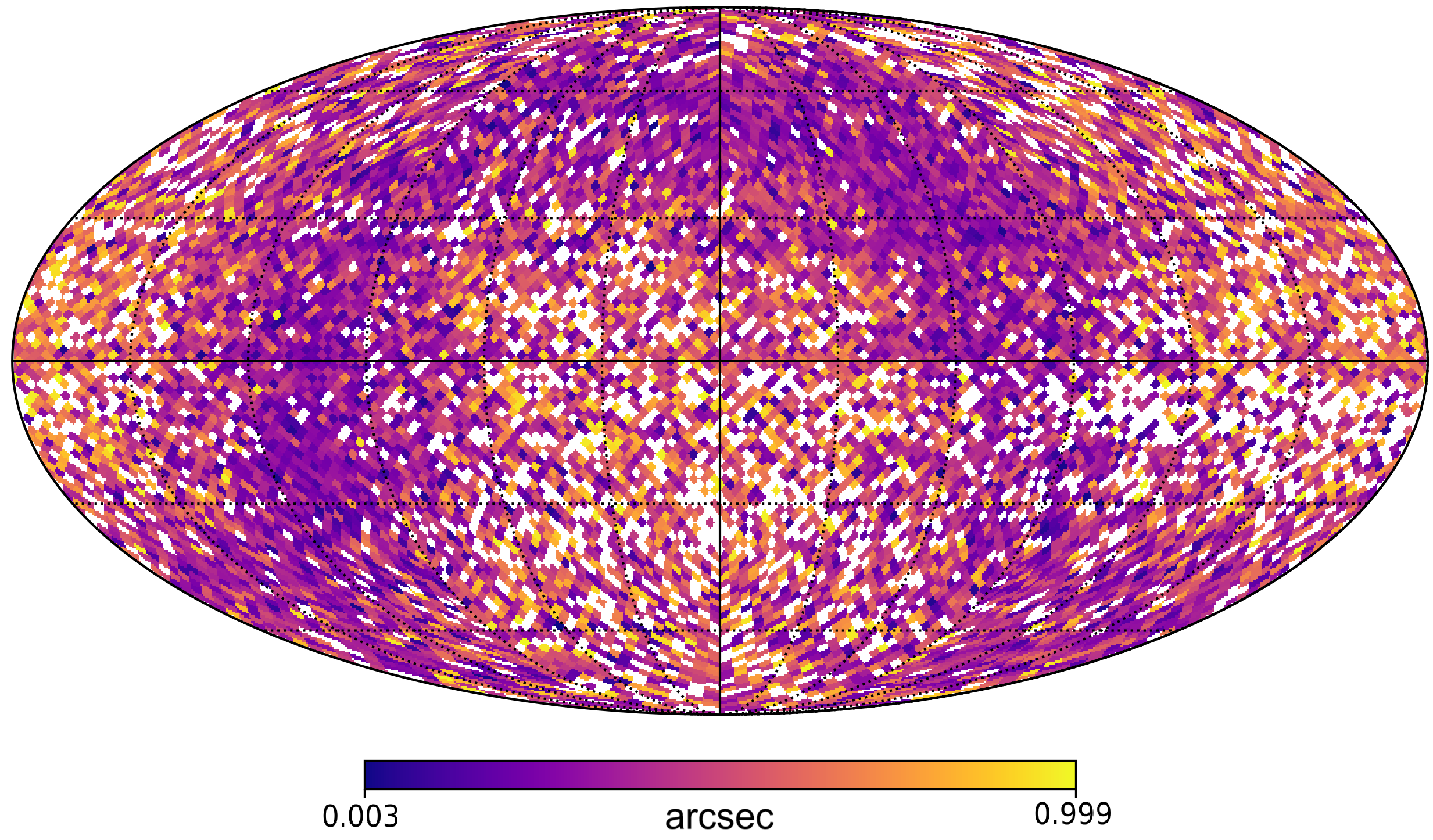}
    \includegraphics[width=0.49\textwidth]{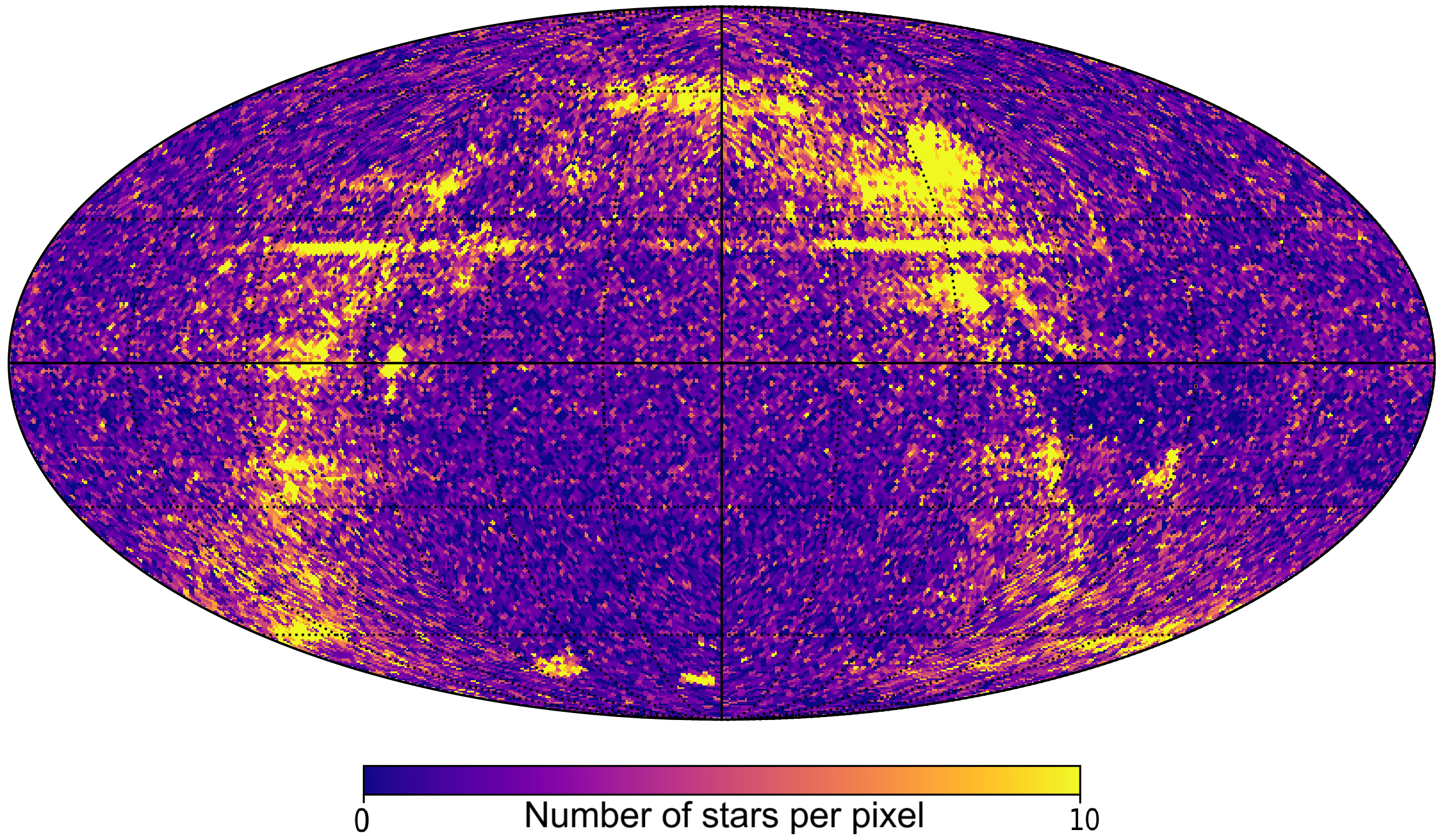}
    \caption{(a) {\edm Average within pixel} distance between identified objects from ILB and the binary star catalog based on Gaia data (NSS or WB), expressed in arcseconds. Pixel size {\edm$\approx$ 3.4 $deg^{2}$}. The image is constructed with equal-area HEALPix pixelization of order 5. 
    Equatorial coordinate system. \\(b) Distribution of stars in the WDS catalog. Yellow color indicates pixels where star density exceeds 10 stars per pixel. Pixel size {\edm$\approx$ 0.8 $deg^{2}$}. The image is constructed with equal-area HEALPix pixelization of order 6. Equatorial coordinate system}
    \label{fig:mean_separation}
\end{figure}



The left panel of Fig.~\ref{fig:mean_separation} shows {{\edm average within pixel} distance between identified objects in equatorial coordinates}. {The division was performed using HEALPix sphere pixelization into equal-area regions ~{https://healpix.sourceforge.io/}, equal to} {\edm$\approx$ 3.4 $deg^{2}$}. The color scale shows the distance magnitude; white color shows pixels where identifications are absent.


{It can be seen that} the distribution of errors across the celestial sphere is significantly non-uniform, with the average identification distance in the Galactic disk being significantly smaller than away from it. On the other hand, white pixels where no identifications are found are concentrated mainly along a "band" crossing the Galactic disk. {This effect is due to selection effects that influenced the creation of the ILB binary star catalog. One of the most important ground-based observation data sources for it is the compilative WDS catalog. The peculiarities of the WDS star distribution across celestial coordinates reflect, in particular, the presence and absence of active binary star observers at certain latitudes.}

\begin{figure}[t]
    \centering
     \includegraphics[width=0.85\linewidth]
         {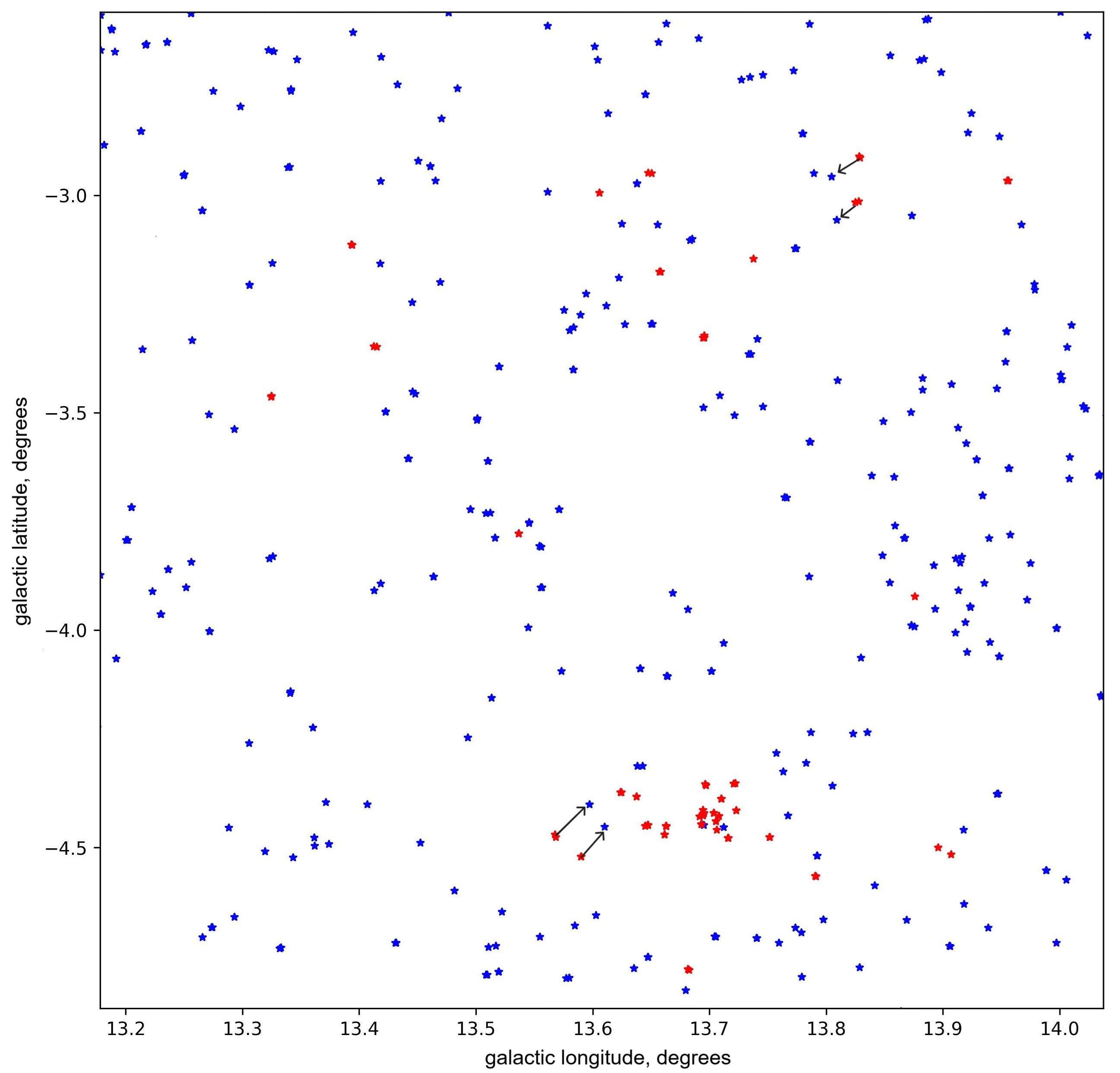}
    \caption{Example of a pixel with no identifications. Red asterisks indicate objects from BDB, blue—Gaia objects. Arrows show some assumed stellar displacements.}
    \label{fig:bad_pix}
\end{figure}

Another probable cause for the absence of identifications is the presence of stellar proper motions (see Fig.~\ref{fig:bad_pix} and Section~\ref{discussion} for details).

\section{Discussion}\label{discussion}

The significant scatter in average distance between identified objects in different sky regions cannot be explained solely by the accuracy of the identified catalogs, since in that case the distributions would show clear boundaries corresponding to regions covered by different catalogs. In this case, a clear correlation is visible between star density and angular distance between identifications, with the average distance between identified objects in dense regions being several times smaller than in regions with few stars. Such a dependence cannot be explained by natural causes, since the difference in coordinate measurement accuracy, both in Gaia and in other sky surveys included in BDB, may depend on stellar field density but has a significantly smaller magnitude (in the case of Gaia—at the level of tenths of a milliarcsecond). The only possible explanation for the observed correlation between star density and identification angular distance may be the presence of a larger number of false identifications in regions where star density is high. Currently, we use the simplest identification method—by the closest object, without accounting for other stellar characteristics. Previously, this method gave good results, since the star density in pre-Gaia identified catalogs was much lower, and the probability that the object closest to the identified star would not be the true identification was small.

Pixels where {\edm identifications} are absent are of particular interest. Analysis showed that in the overwhelming majority of cases, these pixels have no or very few stars from catalogs included in BDB, while Gaia typically has many stars for these pixels. Additionally, identification of objects in binary systems with large proper motion is complicated. As an example, Fig.~\ref{fig:bad_pix} shows stars in one of the pixels where {\edm identifications} are absent. Upon careful examination, it turns out that there are pairs of stars displaced sufficiently strongly that the {\edm cross-identification} algorithm did not work; for some of them, displacements are shown by arrows. Obviously, similar cases exist in other regions as well. As a rule, stars with large proper motion are close to the Sun, and therefore their distribution in galactic coordinates is fairly uniform; accordingly, their fraction in the Galactic disk is smaller, and at high galactic latitudes—larger.
Also, {\edm cross-identification} may fail if there are, conversely, too many candidate stars falling within the specified radius. Such a case is visible in Fig.~\ref{fig:bad_pix} near coordinates [13.73, -4.4].

\begin{figure}[t]
    \centering
     \includegraphics[width=0.85\linewidth]
         {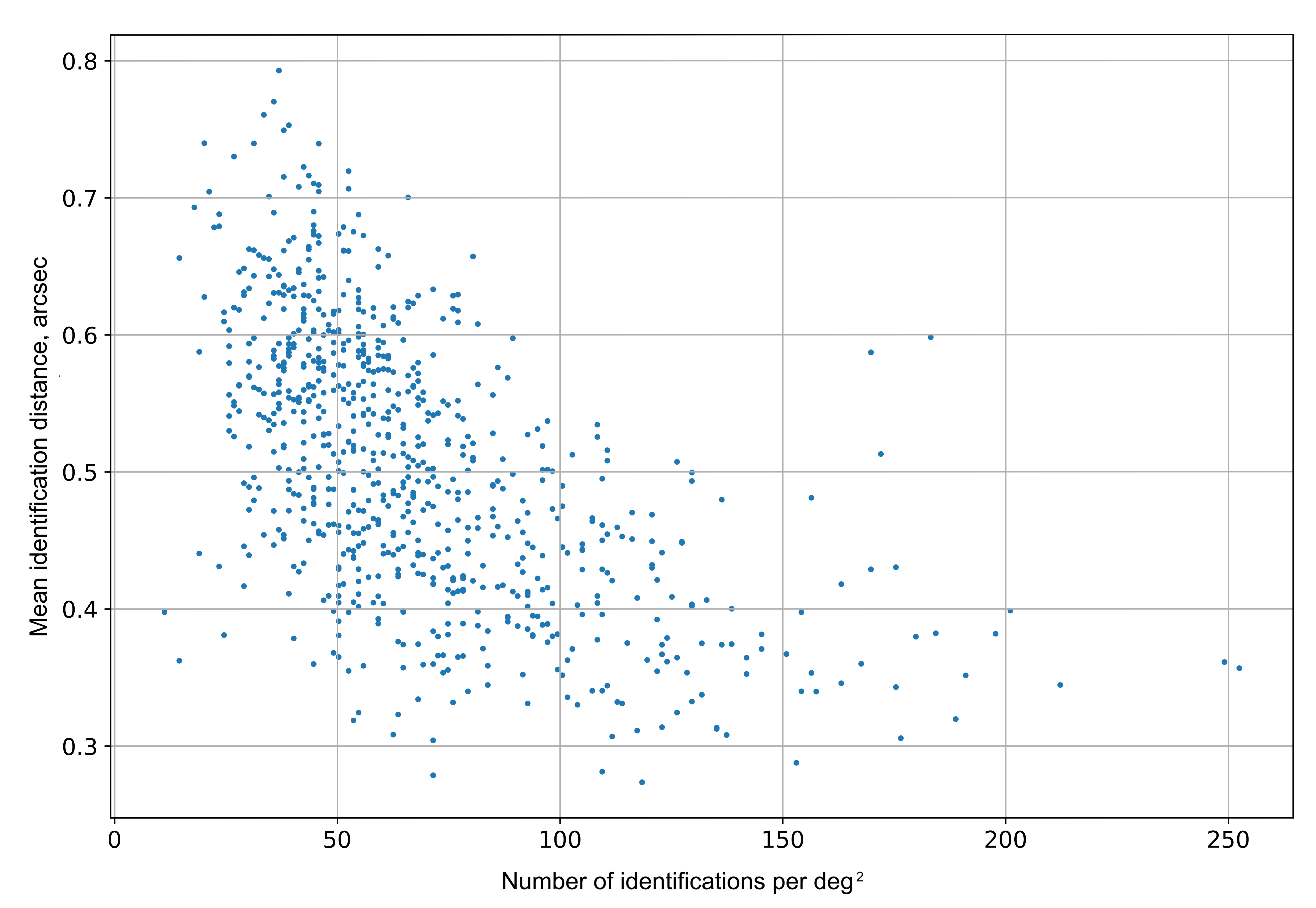}
    \caption{Average distance between identifications within a square degree as a function of the number of stars in this area.}
    \label{fig:sepaartion_number_relation}
\end{figure}

Fig.~\ref{fig:sepaartion_number_relation} shows the dependence of our obtained average distance between identified objects on stellar field density (only identified objects are counted). As can be seen from the figure, there is a very strong, close to linear dependence of average distance on density. Up to a density of about 120 stars per square degree, the average distance falls approximately proportionally to density, after which the dependence reaches a plateau. However, due to the relatively small number of dense areas, there are too few points on the plateau to confidently assert that the dependence actually breaks off.

\section{Conclusion}

In \cite{Malik2024}, we presented the results of cross-identification of stars from binary star catalogs based on Gaia data and data included in the {\edm compilative} ILB catalog. In this work, we investigate the systematic discrepancies found during {\edm cross-identification} between the coordinates of identified stars in ILB and in catalogs based on Gaia data.

Analysis of {\edm cross-identification} results shows that for cross-identification of catalogs, using only the coordinates of identified stars in dense fields is insufficient. It is necessary either to involve additional data about identified stars, such as brightness, or to apply a flexible approach when choosing the {\edm identification} radius. Nevertheless, based on statistical data, it is possible to estimate the probability of false {\edm identification} based on the average distance between stars as a function of celestial coordinates, see~\cite{Kovaleva2025}.

It is shown that accounting for source proper motions is necessary for correct identification. In regions where the average distance between stars is less than the possible displacement due to proper motions, failure to account for them does not introduce significant errors; however, when dealing with catalogs with high surface density of stars, such as Gaia, accounting for proper motions becomes necessary.

{\edm\section{Acknowledgments}
We are grateful to the anonymous reviewers whose constructive comments were very helpful in improving the article.

The authors express sincere gratitude to the "Frontiers of Science" Foundation for supporting the team and this research.}

\normalsize


\bibliographystyle{az}
\bibliography{malik}

\end{document}